\documentclass[a4paper, amsfonts, amssymb, amsmath, preprint, showkeys, nofootinbib, twoside]{revtex4-1}
\usepackage[english]{babel}
\usepackage[utf8]{inputenc}
\usepackage{float}

\usepackage[colorinlistoftodos, color=green!40, prependcaption]{todonotes}
\usepackage{amsthm}
\usepackage{mathtools}
\usepackage{physics}
\usepackage{xcolor}
\usepackage{graphicx}
\usepackage[left=23mm,right=13mm,top=35mm,columnsep=15pt]{geometry} 
\usepackage{adjustbox}
\usepackage{placeins}
\usepackage[T1]{fontenc}
\usepackage{lipsum}
\usepackage{csquotes}
\usepackage[pdftex, pdftitle={Article}, pdfauthor={Author}]{hyperref} 
\begin{document}
\title{The Maximal Gravitational Wave Signal from\\ Asteroid-Mass Primordial Black Hole Mergers\\{At Resonant Microwave Cavities}}

\author{Stefano Profumo}
    \email[Correspondence email address: ]{profumo@ucsc.edu}
    \affiliation{Department of Physics, University of California, Santa Cruz (UCSC),
Santa Cruz, CA 95064, USA}
\affiliation{Santa Cruz Institute for Particle Physics (SCIPP),
Santa Cruz, CA 95064, USA}

\author{Lucas Brown}
    \affiliation{Department of Physics, University of California, Santa Cruz (UCSC),
Santa Cruz, CA 95064, USA}
\affiliation{Santa Cruz Institute for Particle Physics (SCIPP),
Santa Cruz, CA 95064, USA}

\author{Christopher Ewasiuk}
    \affiliation{Department of Physics, University of California, Santa Cruz (UCSC),
Santa Cruz, CA 95064, USA}
\affiliation{Santa Cruz Institute for Particle Physics (SCIPP),
Santa Cruz, CA 95064, USA}

\author{Sean Ricarte}
    \affiliation{Department of Physics, University of California, Santa Cruz (UCSC),
Santa Cruz, CA 95064, USA}
    \affiliation{Department of Physics, University of California, Merced, Merced, CA 95343, USA}

\author{Henry Su}
    \affiliation{Department of Physics, University of Washington, Seattle WA 98195, USA}
\affiliation{Department of Physics, University of Massachusetts, Amherst, MA 01003, USA}

\date{\today} 

\begin{abstract}
Primordial black holes can be the entirety of the dark matter in a broad, approximately five-orders-of-magnitude-wide mass range, the ``asteroid mass range'', between $10^{-16}\ M_{\rm Sun}$ -- where constraints originate from evaporation -- and $10^{-11}\ M_{\rm Sun}$ -- from microlensing. A direct detection in this mass range is very challenging with any known observational or experimental methods. Here we update the calculation of the sight distance for narrow-band detectors such as resonant microwave cavities, and the resulting maximal event rate. We find that the largest detection rates are associated with binaries from non-monochromatic mass functions in early-formed three-body systems. Even in the most optimistic setup, these events are anticipated to be extremely rare.
\end{abstract}


\maketitle
\tableofcontents

\section{Introduction}

Gravitational wave (GW) observations are playing an increasingly central role in the era of multi-messenger astronomy and astrophysics as well as in the search for new physical phenomena. Interferometric detectors like LIGO/Virgo/KAGRA are sensitive to lower-frequency GWs (in the Hz to kHz range), while pulsar timing arrays are sensitive to nano-Hertz frequencies \cite{Maggiore:2007ulw}. At the opposite end of the GW spectrum, high-frequency GWs (HFGWs) require radically novel detection approaches \cite{Dolgov:2011cq}. Broadly, the field of HFGWs, specifically at frequencies in the mega-hertz and above, is a rapidly developing area of research with significant theoretical challenges and technical hurdles (see e.g. Ref.~\cite{Giovannini:2023itq} for a recent overview). 

The experimental landscape of HFGW detectors includes mechanical resonators \cite{Harry:1996gh}, such as resonant spheres, levitated sensor detectors, and bulk acoustic wave devices, broadly operating in the few kHz up to GHz frequencies; devices hinging on the so-called inverse Gertsenshtein effect {(first proposed as a pathway to detect
planetary-mass light black hole mergers with resonant cavities in Ref.~\cite{Herman:2020wao})}, involving the conversion of gravitons into photons \cite{Palessandro:2023tee}, such as superconducting radio-frequency cavities, resonant antennas, and conversion of GWs into electromagnetic waves in the presence of a static magnetic field, as in axion search experiments and axion helioscopes, and resonant LC circuits \cite{Domcke:2022rgu} -- broadly sensitive to frequencies between MHz and GHz; interferometers, such as the HOL experiment, and other techniques, including for instance using the frequency modulation of photons in laser beams or employing superconducting circuits \cite{Bringmann:2023gba}. { The interested Reader is referred to Ref.~\cite{Aggarwal:2020olq} and \cite{Aggarwal:2025noe} for an up to date overview of the experimental landscape.}

A staggering feature of HFGWs is the purported absence of known astrophysical backgrounds, making a detection a potent signal of new physics or of a new class of GW sources. In this study we  focus on one specific source of HFGW: light, nearby primordial black hole (PBH) mergers. PBHs -- BHs of non-stellar origin formed, for instance, from large density fluctuations, on the scale of the cosmic event horizon, in the early universe -- continue to attract growing attention; PBHs could be a large fraction or, depending on their mass, the totality of the Universe's dark matter; be responsible for the production of the dark matter itself, or the matter-antimatter asymmetry in the universe \cite{Morrison:2018xla}; produce observed anomalous gamma-ray and cosmic-ray signals \cite{Picker:2023lup, Korwar:2024ofe}; and provide a window into otherwise utterly secluded dark sectors \cite{Baker:2021btk, Baker:2022rkn}.  The detection of PBH mergers could shed light on early Universe conditions, and on the question of the microscopic nature of the cosmological dark matter. we point the interested Reader to recent reviews on these and related topics, such as Ref.~\cite{Carr:2020xqk, Escriva:2022duf}).

PBH mergers are not the exclusive potential source of HFGWs. Other exotic phenomena include cosmic strings, oscillons, and scenarios involving early Universe dynamics such as inflation (e.g., quintessential inflationary models) \cite{Gehrman, Li}. These sources might produce HFGWs during different early Universe processes, including in certain baryogenesis scenarios \cite{Gehrman}. PBH evaporation could also source very HFGWs \cite{Dolgov:2011cq}, with recently-proposed possible cosmological realizations leading to lower-frequency signals than those traditionally expected at or near the Planck scale \cite{Ireland:2023avg, Ireland:2023zrd}. Some speculative models involve high-energy particle interactions or nuclear processes that might generate detectable GWs at these high frequencies and, broadly, HFGWs offer a new window to test models extending beyond the Standard Model, including those involving extra dimensions or modifications to gravity \cite{Gehrman2}.

HFGW from light PBH mergers has been explored in a number of more or less recent studies. Ref.~\cite{ref1} explores the energy density of relic GWs from PBH mergers; Ref.~\cite{ref9} discusses MHz-GHz GWs related to PBHs and baryogenesis; Ref.~\cite{ref13} considers the stochastic background of high-frequency GWs from early PBH mergers; Ref.~\cite{ref26} explores evaporated PBHs and associated GWs; Ref.~\cite{ref28} examines PBH binary production and GW merger rates. Ref.~\cite{ref3} investigates PBH mergers and induced secondary GWs using inflationary models; Ref.~\cite{ref4} analyzes stochastic background from curvature fluctuations; \cite{ref10} reviews PBH binary formation and merger theories; Ref.~\cite{ref20} offers end-to-end analyses of PBH GW signatures; \cite{ref8} critiques detection prospects for sub-solar PBHs; \cite{ref14} calculates merger rates and evaluates mass distributions related to PBHs; \cite{ref15} discusses inflationary PBHs and resulting GWs; Ref.~\cite{ref18} connects LIGO observations to PBH-induced GWs from cosmic inflation. { Ref.~\cite{Barrau:2023kuv, Barrau:2024kcb} discuss the detection of high-frequency GWs from PBH coalescences and hyperbolic encounters with resonant
cavities.} 

{In the present study, we focus on one promising avenue to search for HFGW from asteroid-mass PBHs: narrow-band resonant cavities, currently used for axion searches, such as ADMX. We show that the sight distance at resonant cavities at asymptotically large PBH masses is mass-independent. In particular, we focus on the conditions leading to the maximal-possible event rate, including} (i) considerable clustering at early times, leading to enhanced early binary formation, (ii) a large over-abundance of dark matter locally, (iii) a non-monochromatic PBH mass function, and (iv) a PBH mass abundance close to the totality of the cosmological dark matter abundance. 


The remainder of this study is structured as follows: in the next section \ref{sec:massfunction} we describe the formalism and constraints on PBH populations of non-monochromatic mass functions; the ensuing sec.~\ref{sec:cavities} outlines and discusses the sensitivity of microwave cavities to HFGWs specifically originating from PBH mergers, including with non-monochromatic mass functions;  sec.~\ref{sec:rates} then focuses on the most significant pathway leading to the largest merger event rate, and on the corresponding rate of detectable events at an ADMX-like microwave cavity -- we leave it to the Appendix to present a comparison to other possible pathways, leading to considerably subdominant event rates; the final sec.~\ref{sec:conclusions} then presents our discussion and conclusions. 

\section{Constraints for a generic mass function}\label{sec:massfunction}
We consider a mass distribution of PBHs (the so-called ``mass function''), the mass-weighted differential number of PBHs, of the standard form
\begin{equation}
\psi(m)\equiv \frac{1}{m}\frac{dn_{\rm PBH}(m)}{dm},
\end{equation} 
normalized so that the cosmological mass density of PBHs $\rho_{\rm PBH}$ relative to the cosmological dark matter abundance $\rho_{\rm DM}$ 
\begin{equation}
f_{\rm PBH}=\int \psi(m)dm.
\end{equation}
We refer to monochromatic mass functions, corresponding to the form
\begin{equation}\label{eq:mono}
\psi_0(m)=f_{\rm PBH}(m_0)\delta(m-m_0),
\end{equation}
and to ``dichromatic'' mass functions,
\begin{equation}\label{eq:dich}
    \psi_{\rm DC}(m)=f_1\delta(m-m_1)+f_2\delta(m-m_2).
\end{equation}
The latter is especially well motivated in the present context, as previous studies have numerically shown that the structure of mass functions that maximize black hole merger rates are generally rather close to the schematic form in Eq.~(\ref{eq:dich}) (see e.g. Ref.~\cite{Lehmann:2020bby}).

Note that translating from the customarily shown constraints on monochromatic mass functions, representing the largest-possible value of $f^{\rm max}_{\rm PBH}(m_0)$ for Eq.~(\ref{eq:mono}) above compatible with experimental and observational constraints, to a generic mass function $\psi(m)$ amounts to requiring that \cite{Carr:2017jsz}
\begin{equation}
\int\frac{\psi(m)}{f^{\rm max}_{\rm PBH}(m)}dm\le1.
\end{equation}
Hence, the resulting constraint on $f_1$ and $f_2$ for a monochromatic mass function is (dropping the superscript ``max'' from now on):
\begin{equation}
    \frac{f_1}{f_{\rm PBH}(m_1)}+\frac{f_2}{f_{\rm PBH}(m_2)}\le1.
\end{equation}
For a given $f_1$, the value of $f_2$ that maximizes the mass-density of PBHs is then fixed by
\begin{equation}\label{eq:f2max}
    f_2={\rm Max}\left(f_{\rm PBH}(m_2)\left(1- \frac{f_1}{f_{\rm PBH}(m_1)}\right),0\right)={\rm Max}\left(\frac{f_{\rm PBH}(m_2)}{f_{\rm PBH}(m_1)}\left(f_{\rm PBH}(m_1)-f_1\right),0\right).
\end{equation}

\section{Microwave cavities: sensitivity}\label{sec:cavities}
Resonant cavities -- first proposed to search for axions \cite{Sikivie:1983ip} -- are sensitive to high-frequency GWs: a GW passing in a cavity endowed with a static magnetic field sources an effective electromagnetic current that, in turn, generates an electromagnetic field at the same frequency of the GW \cite{Franciolini207, Franciolini208, Franciolini209, Gatti:2024mde}. Currently-operating and proposed detectors include ADMX \cite{ADMX:2018ogs, ADMX:2019uok, ADMX:2021nhd}, HAYSTAC \cite{HAYSTAC:2018rwy}, CAPP \cite{Lee:2020cfj}, and ORGAN \cite{McAllister:2017lkb}. Here, for definiteness, we consider an ADMX-type detector for our estimates. In particular, we adopt the following estimate for ADMX's sensitivity to the GW amplitude \cite{Franciolini:2022htd}:
\begin{eqnarray}\label{eq:cavity}
    h_{\rm sens}(\nu)&=&3\times 10^{-22}\left(\frac{0.1}{\eta_n}\right)^{}\left(\frac{8\ {\rm T}}{|\vec{B}|}\right)^{}\left(\frac{0.1\ {\rm m}^3}{V_{\rm cav}}\right)^{5/6}\left(\frac{10^5}{Q}\right)^{1/2}\\
    &&\left(\frac{T_{\rm sys}}{1\ {\rm K}}\right)^{1/2}\left(\frac{1\ {\rm GHz}}{\nu}\right)^{3/2}\left(\frac{\Delta \nu}{10\ {\rm kHz}}\right)^{1/4}\left(\frac{1\ {\rm min}}{\Delta t}\right)^{1/4}.\nonumber
\end{eqnarray}
In the equation above, $\eta_n$ is the cavity's effective GW coupling; 
Note that in the most-optimal possible case, i.e. when the azimuthal direction of the relevant cavity mode resonates with the spin structure of the gravitational field, the cavity’s effective coupling coefficient $\eta_n$ is ${\cal O}(0.1)$~\cite{Franciolini207}. This includes the $TM_{010}$ and $TM_{020}$ modes in use for axion experiments like ADMX. It was also found in ~\cite{Gatti:2024mde} that $\eta_n \approx 0.14$ for the $TM_{012}$ mode. Note that while \cite{Franciolini:2022htd} presents the coupling coefficient as ${\cal O}(1)$, the authors used $\eta_n = 0.1$ in their calculations for the strain sensitivities of ADMX and SQMS. We use the same here, albeit with future optimization this may actually be an under-estimate.

Also in Eq. (\ref{eq:cavity}), $|\vec{B}|$ is the cavity's magnetic field; $Q$ is the quality factor of the cavity; $\nu$ the operating resonant frequency; and $\Delta \nu\sim \nu/Q$ the bandwidth. The other parameters are the cavity's volume, operating temperature, and acquisition observation time. Unless otherwise specified, we will utilize the pivot values in Eq.~(\ref{eq:cavity}) as our reference values for the sensitivity of an ADMX-like microwave cavity. 


As customary, we define the chirp mass of the binary as 
\begin{equation}
    m_c\equiv\frac{(m_1m_2)^{3/5}}{(m_1+m_2)^{1/5}}.
\end{equation}
{The gravitational wave amplitude for a merger of a binary with chirp mass $m_c$ at a distance $d$ reads (see e.g. Eq.(4.3) and (4.29) of Ref.~\cite{Maggiore:2007ulw})
\begin{equation}\label{eq:strain}
    h(m_c,d,\nu)=\frac{4}{d}\left(\frac{Gm_c}{c^2}\right)^{5/3}\left(\frac{\pi \nu}{c}\right)^{2/3},
\end{equation}
at frequency $\nu$. The time derivative of the frequency reads (Eq. (4.18) in Ref.~\cite{Maggiore:2007ulw})
\begin{equation}
    \dot \nu=\frac{96}{5}\pi^{8/3}\left(\frac{Gm_c}{c^3}\right)^{5/3}\nu^{11/3}.
\end{equation}


The signal in the cavity depends on whether the cavity is fully ``rung-up'' or not; in turn, this depends on comparing the so-called {\em ring-up time} of the cavity, $t_r=Q/\nu_0=10^{-4}\ {\rm sec}$, the latter value referring to the specific case of ADMX, with the time spent at frequency $\nu$, given by
\begin{equation}
    t_g(M,\nu)=\int_{\nu_{\rm min}}^{\nu_{\rm max}} \frac{d\nu}{\dot \nu},
\end{equation}
where $\nu_{\rm max,min}=\nu_0\pm\nu_0/Q$, with $\nu_0=1$ GHz the resonant frequency of the cavity. Carrying out the integral, and noting that $Q^2-1\simeq Q^2$ for a narrow-band cavity, we find
\begin{equation}
    t_g(M,\nu)=\frac{5}{48}\frac{1}{Q}(\pi\nu_0)^{-8/3}\left(\frac{GM}{c^3}\right)^{-5/3}.
\end{equation}
Note that in fact one should use, above, the Lorentzian suppression of the cavity's response around the resonance frequency, and the proper integral is
\begin{equation}
    t_{g^\prime}=\int \frac{d\nu}{\dot \nu}\frac{1}{1+\left(\frac{\nu-\nu_0}{\nu_0/Q}\right)^2};
\end{equation}
using the expression above, however, we find that the result is very close, to a few percent, to the estimate that neglects the Lorentzian factor.

Note that the expressions above do not follow the behavior expected for broad-band detectors, 
\begin{equation}
    N^{\rm broad}_{\rm cycles}=\int_{t_{\rm min}}^{t_{\rm max}} \nu dt=\int_{\nu_{\rm min}}^{\nu_{\rm max}}d\nu\frac{\nu}{\dot \nu}\simeq \frac{\nu^2}{\dot \nu},
\end{equation}
with the characteristic time spent at a frequency $\nu$ is
\begin{equation}
    t^{\rm broad}_g(m_c,\nu)=\frac{\nu}{\dot \nu},
\end{equation}
as quoted e.g. in Ref.~\cite{Franciolini:2022htd}, Eq.~(2.33), and Ref.~\cite{Gatti:2024mde}, Eq.~(12).

The ring-up condition for a resonant cavity can also be expressed in terms of the quality factor and the number of cycles the signal spends at the resonant frequency. The latter, i.e. the number of cycles within the cavity's bandwidth is 
\begin{equation}
    N_c=\int_{\nu_{\rm min}}^{\nu_{\rm max}} \frac{\nu}{\dot\nu}d\nu\simeq \frac{5}{48}\frac{\nu_0}{(\pi\nu_0)^{8/3}Q}\left(\frac{GM}{c^3}\right)^{-5/3},
\end{equation}
and thus the condition for the cavity to be rung up is equivalently expressed as $N_c=Q$.

The signal in the cavity follows from he ring-up equation for resonant cavities
\begin{equation}
h_{\rm sig}(m_c,d,\nu)=h(m_c,d,\nu)\ {\rm min}\big[1-{\rm exp}\left(N_c(m_c,\nu)/Q\right)].
\end{equation}

We define the sight distance, or distance sensitivity, for a chirp mass $m_c$ at a frequency $\nu$ as the distance at which 
\begin{equation}\label{eq:sens}
    h(m_c,d_{\rm sens},\nu)=h_{\rm sens}(\nu).
\end{equation}

Note that for $N_c/Q\ll1$, $1-{\rm exp}\left(N_c/Q\right)\simeq N_c/Q$; therefore, the signal in the cavity goes as
\begin{equation}\label{eq:straincavity}
    {\rm lim}_{N_c\ll Q}h_{\rm sig}(m_c,d,\nu)=\frac{4}{d}\left(\frac{Gm_c}{c^2}\right)^{5/3}\left(\frac{\pi \nu}{c}\right)^{2/3}\frac{5}{48}\frac{\nu}{(\pi\nu)^{8/3}Q}\left(\frac{Gm_c}{c^3}\right)^{-5/3}=\frac{5}{12\pi^2 d}\frac{c}{\nu_0 Q},
\end{equation}
and is thus independent of mass; the distance sight then simply goes as 
\begin{equation}\label{eq:dsens}
    d_{\rm sens}\sim0.03\ {\rm AU}\left(\frac{\Delta\nu}{10\ {\rm kHz}}\right)^{1/4}=0.03\ {\rm AU}\left(\frac{\nu}{{\rm 1\ GHz}}\frac{10^5}{Q}\right)^{1/4}.
\end{equation}
45
}

\begin{figure}[!t]
    \centering
     \mbox{\includegraphics[height=6.3cm]{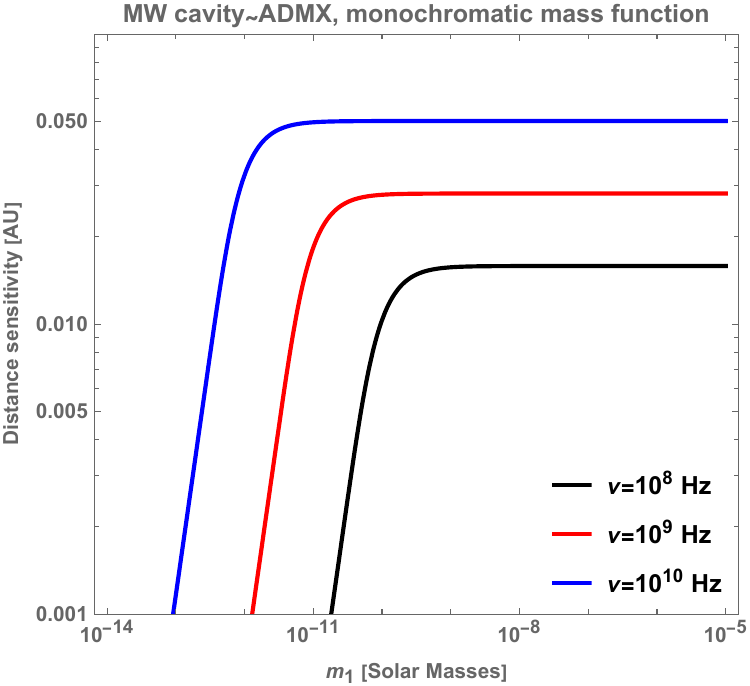}\qquad \quad\quad\includegraphics[height=6.3cm]{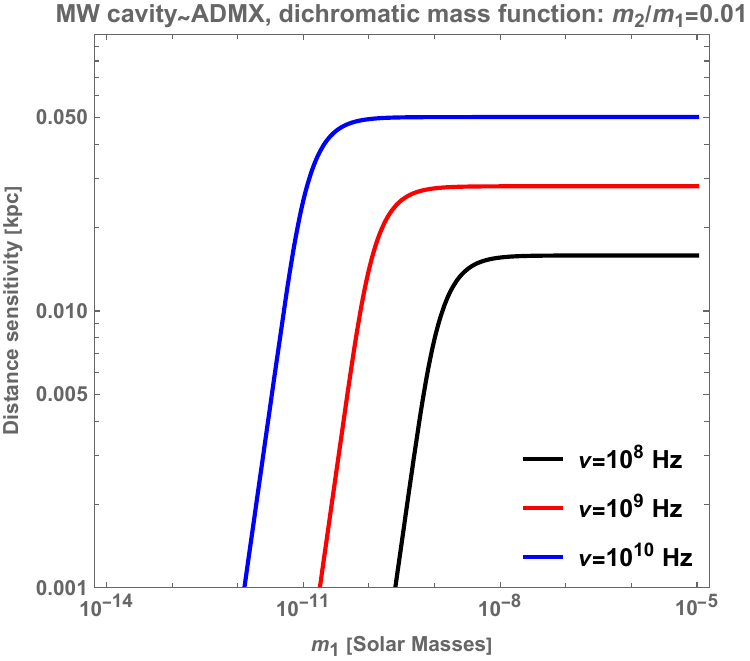}
    }\\[1cm]
    \mbox{\includegraphics[height=6.3cm]{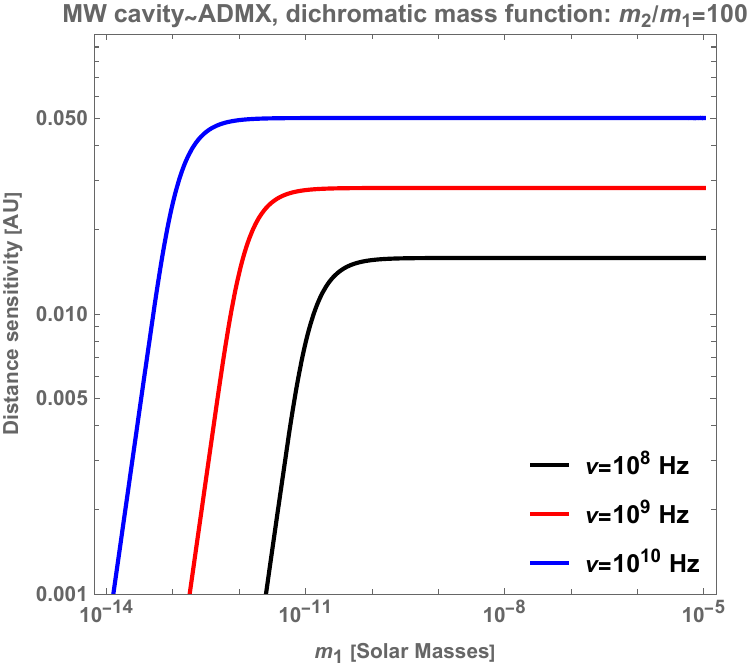}\qquad \qquad \includegraphics[height=6.4cm]{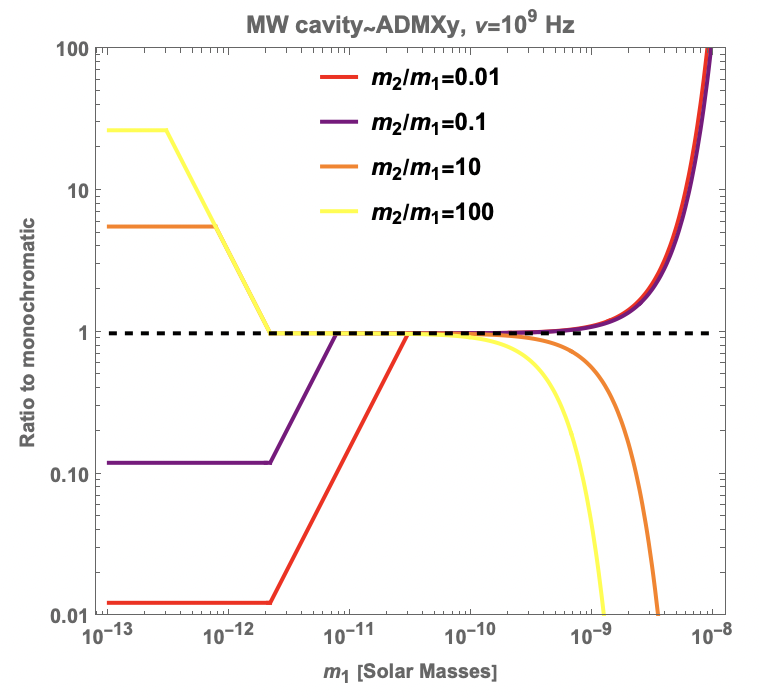}
    }
    \caption{Distance sensitivity comparison for a microwave cavity with specifications close to ADMX's for three different frequencies, $\nu=10^8,\ 10^9\ {\rm and}\ 10^{10}$ Hz. In the top left corner we assume a monochromatic mass function, in the top right that $m_2/m_1=10^{-2}$, in the bottom left that $m_2/m_1=10^{2}$; in the bottom right we instead show the ratio, for $\nu=10^9$ Hz, of the distance sensitivity for a given mass ratio $m_2/m_1$ as in the legend, divided by that for the monochromatic case, $m_1=m_2$}
    \label{fig:ADMXsens}
\end{figure}

We show the ADMX distance sensitivity at a frequency $\nu$ for mergers involving two masses $m_1$ and $m_2$ in Fig.~\ref{fig:ADMXsens}. The top left panel shows the monochromatic mass function case, $m_1=m_2$ for three different frequencies, $\nu=10^8,\ 10^9\ {\rm and}\ 10^{10}$ Hz, while the top right panel assumes a di-chromatic mass function of mass ratio $m_1/m_2=0.01$ and the bottom left panel of $m_1/m_2=0.01$, again for the same three frequencies. Note that the shape of the curves as a function of mass reflects the mass dependence of the sensitivity function $d_{\rm sens}(m_1,m_2,\nu)$; {note, in particular, the chirp-mass-independent sight distance at large masses ($N_c/Q\ll1$) discussed above; also note that ADMX's sensitivity, corresponding, approximately, to the red lines, is typically around 0.03 AU only.}

In the bottom right panel of Fig.~\ref{fig:ADMXsens} we show, for $\nu=10^9$ Hz, the ratio of the distance sensitivity for a given mass ratio $m_2/m_1$ to that corresponding to a monochromatic mass function ($m_1=m_2$), as a function of $m_1$. The plot illustrates how mass function ratios, $m_2/m_1<1$, can be detected at a greater distance than the monochromatic case, due to presence of lighter binaries that can ring up the cavity for larger values of $m_1$; { similarly, for high mass ratios, $m_2/m_1>1$ the largest merger ratios occur at smaller $m_1$, again, that is, when one of the masses in the binary is sufficiently small. We note that in any case {\it a non-monochromatic (here, a dichromatic) mass function generically enables larger distance sensitivities than the monochromatic case}, for sufficiently spaced-out masses.}






\section{Rate of detectable events from light black hole mergers}\label{sec:rates}

In what follows we are concerned with the calculation of the rate of detectable events from light PBH mergers, principally with a monochromatic or dichromatic mass function, at a microwave resonant cavity such as ADMX.

For a generic mass function $\psi(m)$, the {\em rate} of visible mergers at frequency $\nu$, for a detector whose critical distance sensitivity is given by the function $d_{\rm sens}(m_1,m_2,\nu)$ in Eq.~(\ref{eq:dsens}), reads
\begin{equation}
    R[\psi](f)=\int^{\infty}_{m_{\rm min}} dm_1\int^{\infty}_{m_{\rm min}} dm_2\ \frac{dR[\psi](f)}{dm_1dm_2}\frac{4\pi}{3} d_{\rm sens}^3(m_1,m_2,f),
\end{equation}
where $m_{\rm min}$ corresponds to black holes long-lived enough not to have expired yet and $R$ is the merger rate for the given mass function and frequency. 

We demonstrate in the Appendix that the largest differential merger rate today corresponds to the early three-body binary formation pathway ($E3$ in what follows), discussed in detail in Ref.~ \cite{Raidal:2017mfl}
 (see their Eq. (2.13)):
 \begin{eqnarray}
    \nonumber\frac{dR_{E3}}{dm_1dm_2dm_3}&=&\frac{9}{296\pi}\frac{1}{\tilde\tau}\left(\frac{t_0}{\tilde\tau}\right)^{-34/37}\left(\Gamma\Bigg[\frac{58}{37},\tilde N \left(\frac{t_0}{\tilde\tau}\right)^{3/16}\Bigg]-\Gamma\Bigg[\frac{58}{37},\tilde N \left(\frac{t_0}{\tilde\tau}\right)^{-1/7}\Bigg]\right)\\
    && \times \tilde x^{-3}\delta_{\rm dc}^{-1}\tilde N^{53/37}\bar m^3\frac{\psi(m_1)}{m_1}\frac{\psi(m_2)}{m_2}\frac{\psi(m_3)}{m_3}.
\end{eqnarray}
In the equation above (restoring factors of $c$ suppressed in the original reference), 
\begin{equation}
    \tilde\tau\equiv {c^5} \frac{384}{85}\frac{\alpha^4\beta^7 a_{\rm eq}^4m_3^7\tilde x^4}{G^3\eta M^{10}},
\end{equation}
where we will hereafter assume $\alpha=\beta=1$,  
\begin{equation}
    \eta\equiv\frac{m_1m_2}{M^2},\qquad M\equiv m_1+m_2,
\end{equation}
and
\begin{equation}\label{eq:tildex}
    \tilde x^3\equiv\frac{3}{4\pi}\frac{M}{a_{\rm eq}^3\rho_{\rm eq}}.
\end{equation}
Finally, 
\begin{equation}
    \tilde N\equiv \delta_{\rm dc}\Omega_{\rm DM,eq}\frac{M}{\bar m},\ \ {\rm and}\ \ \bar m\equiv\left(\int dm\frac{\psi(m)}{m}\right)^{-1}.
\end{equation}
In the expressions above, the matter-radiation equality corresponds to the energy density relative to the critical density, and expansion factor 
\begin{equation}
    \Omega_{\rm eq}=0.4,\ \ {\rm with}\ \ a_{\rm eq}\simeq1/3400.
\end{equation}
A crucial input to the expressions above is the local density contrast $\delta_{\rm dc}$ at the decoupling redshift corresponding to 
\begin{equation}
a_{\rm dc}\approx a_{\rm eq}\left(\frac{x}{\tilde x}\right)^3,
\end{equation}
where $x\equiv r/a$ is the comoving distance. Assuming that the PBH two-point function $\xi(x)$ is constant at comoving distances smaller than $\tilde x$ specified above in Eq.~(\ref{eq:tildex}), i.e.
\begin{equation}
1+\xi(x)\approx \delta_{\rm dc}\ \ {\rm for}\ \ x<\tilde x,
\end{equation}
 strong clustering corresponds to $\delta_{\rm dc}\gg1$ \cite{Raidal:2017mfl}. Following Ref.~\cite{Franciolini:2022htd} and \cite{Pujolas:2021yaw}, in order to assess the maximal-possible merger rate we additionally consider a large local enhancement factor $\delta_{\rm local}\sim 2\times 10^5$, linearly impacting the merger rate $R\to\delta_{\rm local}R$,  reflecting the fact that the merger rate is sensitive to the {\it local} dark matter density -- i.e. assuming that the binary density follows the dark matter density.


We note that the expression above can be {\em maximized} by expanding the Gamma function in the asymptotically large $\delta_{\rm dc}$ limit, by using the asymptotic expansion for the incomplete Gamma function
\begin{equation}
    \Gamma[a,z]\simeq e^{-z}z^{a-1} \qquad z\to \infty.
\end{equation}
Neglecting the second Gamma function, which is very suppressed compared to the first one for large values of the second argument, one obtains
\begin{eqnarray}
    \nonumber\frac{dR_{E3}}{dm_1dm_2dm_3}&=&\frac{9}{296\pi}\frac{1}{\tilde\tau}\left(\frac{t_0}{\tilde\tau}\right)^{-13/16}\left(\exp[-\tilde N \left(\frac{t_0}{\tilde\tau}\right)^{3/16}]\right)\tilde N^2 \\
    && \times \tilde x^{-3}\delta_{\rm dc}^{-1}\bar m^3\frac{\psi(m_1)}{m_1}\frac{\psi(m_2)}{m_2}\frac{\psi(m_3)}{m_3}.
\end{eqnarray}
We will not use the expression above, which is, however, relevant to interpret our numerical results below.

{ Note that important physical effects can alter the quantitative values for the merger rates outlined above here. For instance, Ref.~\cite{Raidal:2018bbj} finds, with dedicated numerical simulations, that the rates for
early binaries can be importantly reduced when one excludes binaries
ending up in a PBH cluster induced by Poisson fluctuations, a suppression factor
that can be lower than $10^{-2}$ for large PBH fraction. We elaborate on this point in the Appendix, showing that even then the early 3-body pathway would dominate over all other pathways, even if the problem
is amplified in scenarios with initial clustering, whereby regular interactions in clusters would reduce the predicted rates. Additionally, we note here that the PBH halos induced by a strong initial clustering
could be dynamically instable and expand, as this is the case for PBH clusters
induced by Poisson fluctuations, albeit it is unclear if this effect would impact the
early 3-body merger rates. We give a detailed discussion of these effects, as well as the issue of how clustering of PBHs is constrained by direct methods such as microlensing and GWs, in Appendixes B and C below.

}

\begin{figure}[!t]
    \centering
    \mbox{\includegraphics[height=7.5cm]{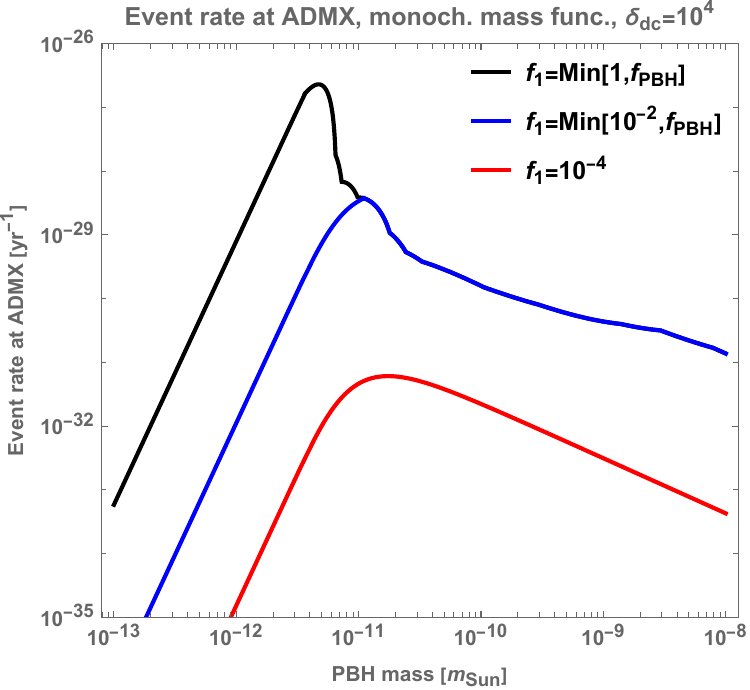}\ \includegraphics[height=7.5cm]{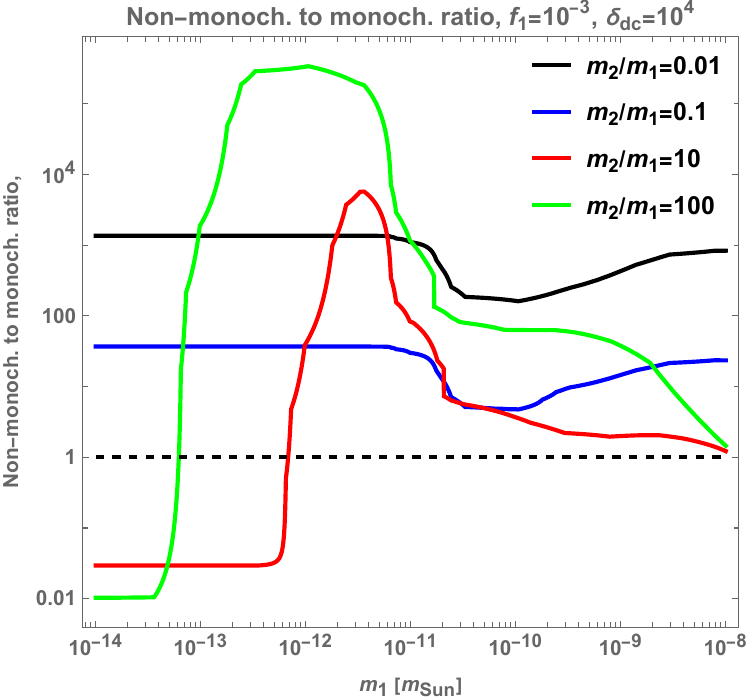}
    }
    \caption{Left: rates (in events per year) for a monochromatic mass function with a fixed mass fraction $f_1$ (if $f_1>f_{\rm PBH}$, i.e. when the assumed mass fraction violates observational limits, we set the mass fraction to its maximal possible value $f=f_{\rm PBH}$). Right: the ratio, as a function of $m_1$, of event rates for dichromatic mass functions with a given $0.01\le m_2/m_1\le 100$, to the monochromatic case, for $f_1=10^{-3}$  and for $\delta_{\rm dc}=10^4$}
    \label{fig:rates}
\end{figure}
Fig.~\ref{fig:rates}, left, shows the merger rates for a monochromatic mass function with a fixed mass fraction $f$ (if $f>f_{\rm PBH}$, i.e. when the assumed mass fraction violates observational limits, we set the mass fraction to its maximal possible value $f=f_{\rm PBH}$). The peak occurs at masses where (i) the constraints on $f_{\rm PBH}$ allow PBH to be the entirety of the dark matter, and (ii) the distance sensitivity plateaus at its maximal value, for the frequency under consideration.

The right panel of fig.~\ref{fig:rates} shows the ratio, as a function of $m_1$ of event rates for dichromatic mass functions with a given { mass ratio} $0.01\le m_2/m_1\le 100$, to the monochromatic case, for $f_1=10^{-3}$  and for $\delta_{\rm dc}=10^4$ (this latter parameter is largely uninfluential here). { The plot bears out what commented on for the bottom, right panel of fig.~\ref{fig:ADMXsens}: large mass ratios produce enhanced rates for smaller $m_1$, and smaller mass ratios for larger $m_1$, and the event rates are generally larger for dichromatic mass functions than for monochromatic mass functions. The non-trivial features in the figure casuing  non-monotonic behaviors are due to a non-trivial combination of the  sensitivity distance and constraints on the abundance of PBHs at a given mass. }

\begin{figure}[!t]
    \centering
    \mbox{\includegraphics[height=7.8cm]{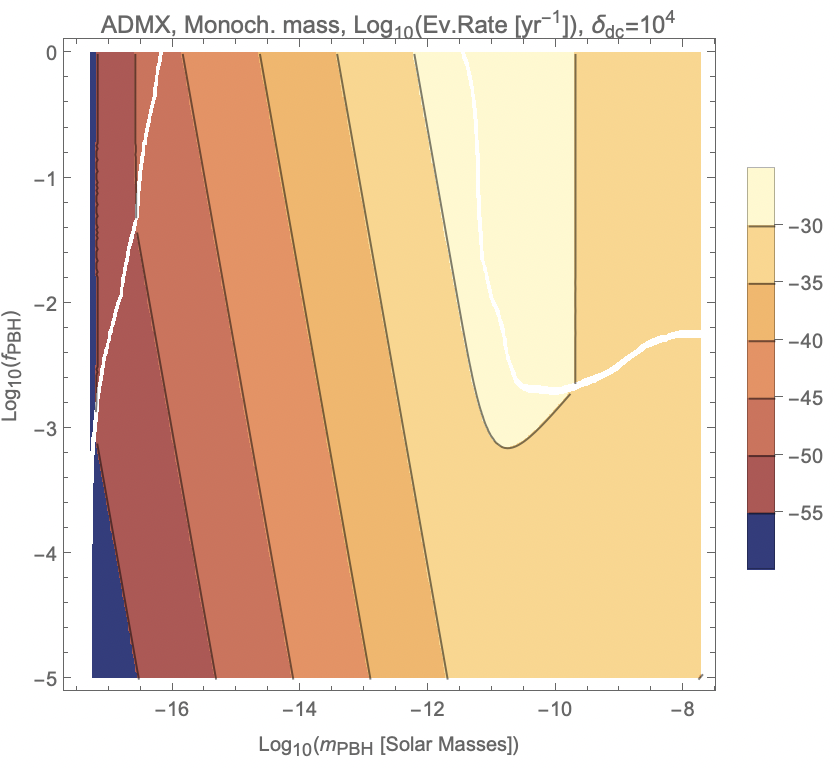}\quad \includegraphics[height=7.8cm]{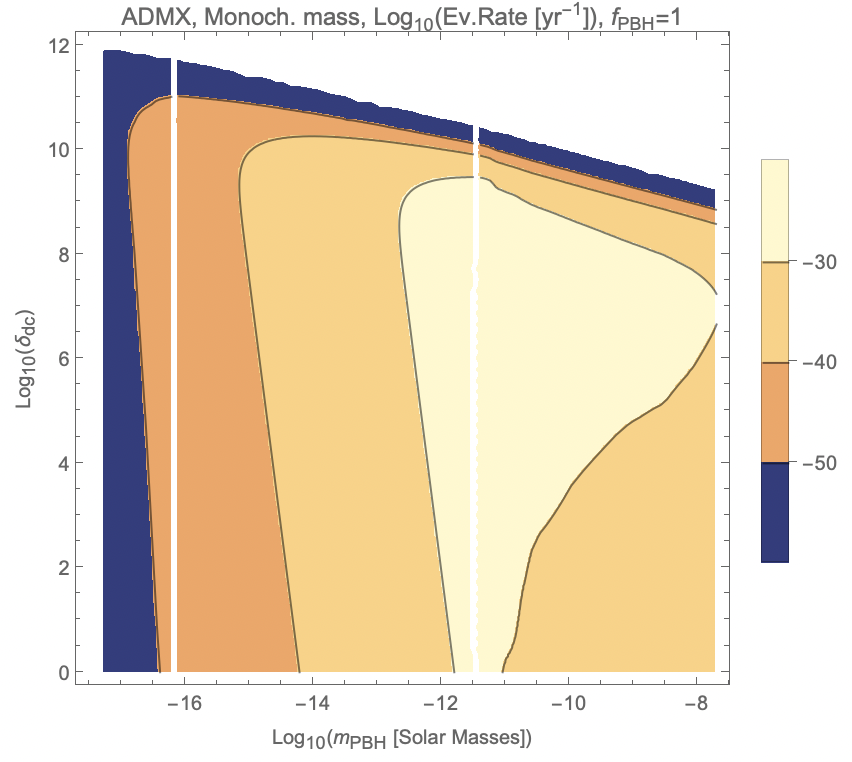} 
    }
    \caption{Contours of constant Log10 of the event rate per year for a putative ADMX-like cavity for a monochromatic mass function varying the binary merger mass $m_{\rm PBH}$ and $f_{\rm 1}$ with $\delta_{\rm dc}=10^4$ (left), and varying instead $\delta_{\rm dc}$ with $f={\rm min}\left(f_{\rm 1},1\right)$}
    \label{fig:ratescntr}
\end{figure}
The contour plots in fig.~\ref{fig:ratescntr} show, on a Log10 scale, the event rate per year for a putative ADMX-like cavity for a monochromatic mass function varying the binary merger mass and $f_{\rm PBH}$ (left) with $\delta_{\rm dc}=10^4$ (left), and varying instead $\delta_{\rm dc}$ with $f={\rm min}\left(f_{\rm PBH},1)\right)$. The sharp decrease on the  left side of the  plots is related to increasingly tight constraints on the abundance of PBH at those masses; we also note that the region with { the largest event rates} extends over a relatively broad range of masses, $10^{-12}\lesssim m_{\rm PBH}/m_{\rm Sun}\lesssim 10^{-8}$, and down to PBH densities of about 0.1\% of the dark matter density; additionally, the left panel highlights the relatively weak dependence of the event rate on $\delta_{\rm dc}$ outside of the exponential suppression at $\delta_{\rm dc}\gtrsim 10^9$, noted already e.g. in  \cite{Franciolini:2022htd}.


\begin{figure}[!t]
    \centering
    \mbox{\includegraphics[height=7.5cm]{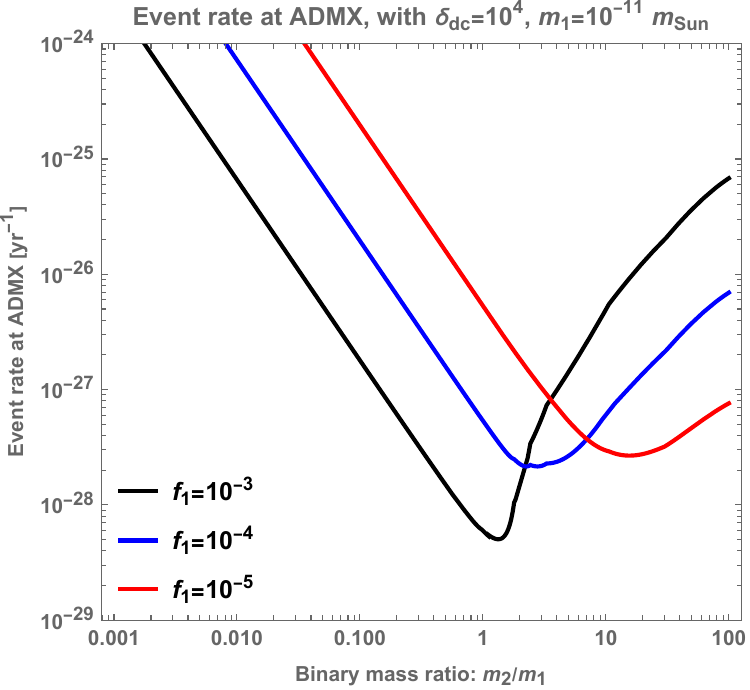}\qquad \includegraphics[height=7.5cm]{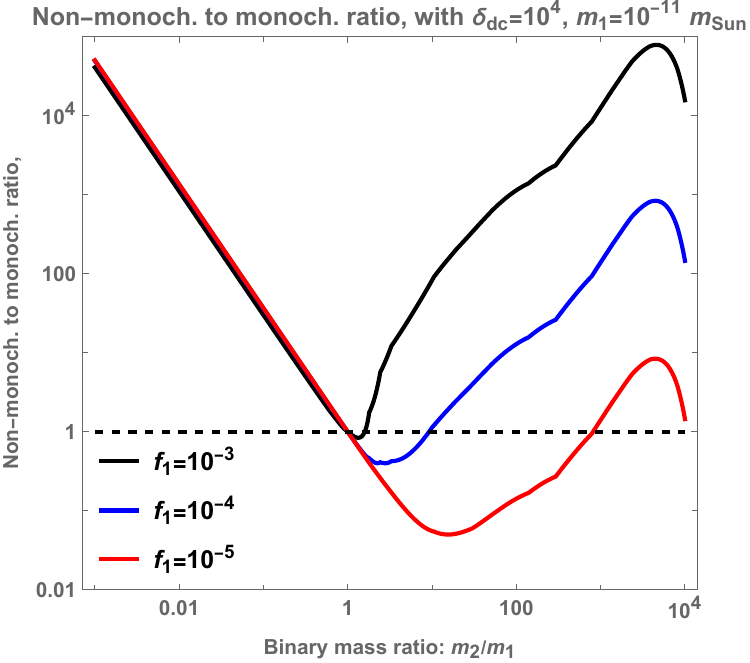}
    }
    \caption{Left: Event rate at an ADMX-like cavity for different $f_1=10^{-3},\ 10^{-4},\ 10^{-5}$ as a function of the binary mass ratio $m_2/m_1$, with $m_1=10^{-11}\ m_{\rm Sun}$, for $\delta_{\rm dc}=10^4$. Right: same as in the left panel, but as a ratio to the monochromatic $m_2=m_1$ rate}
    \label{fig:rates2}
\end{figure}
Fig.~\ref{fig:rates2} relaxes the assumption of a monochromatic mass function, and considers, for three different values of the ``weight'' (relative abundance) $f_1$ of the population with mass $m_1$ (see Eq.(\ref{eq:dich})), the event rate as a function of the mass ratio $m_2/m_1$. Note that here and hereafter we assume the maximal $f_2$, fixed by Eq.~(\ref{eq:f2max}) for $f_1\leq_{\rm PBH}(m_1)$, otherwise the curves are interrupted as the abundance exceeds limits. In the figure, we assume $m_1=10^{-11}\ m_{\rm Sun}$, for $\delta_{\rm dc}=10^4$. The right panel shows the same, but as a ratio to the monochromatic $m_2=m_1$ case, and highlights how a non-monochromatic mass ratio predicts, for $m_2$ between 0.05$m_1$ and $0.9m_1$, a larger event rate than the monochromatic case. 

\begin{figure}[!t]
    \centering
    \mbox{\includegraphics[height=7.5cm]{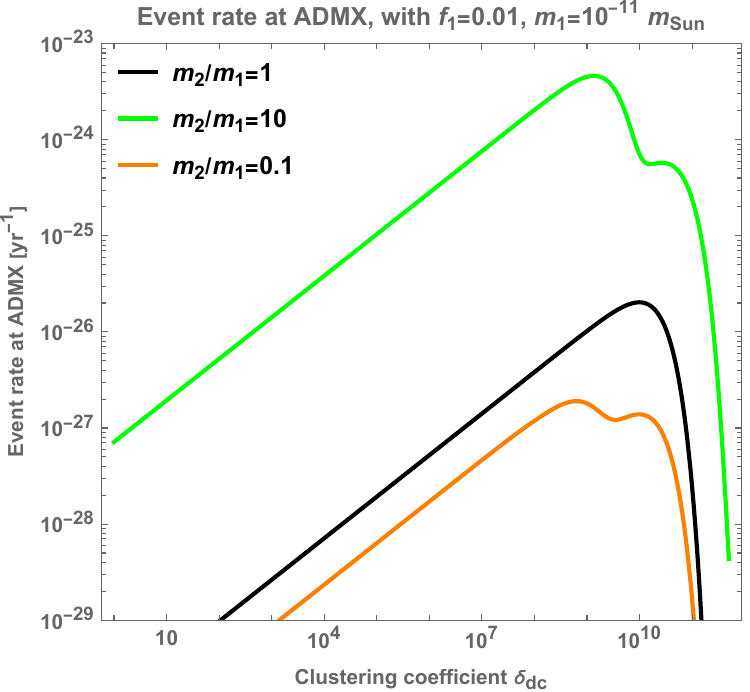}\qquad \includegraphics[height=7.5cm]{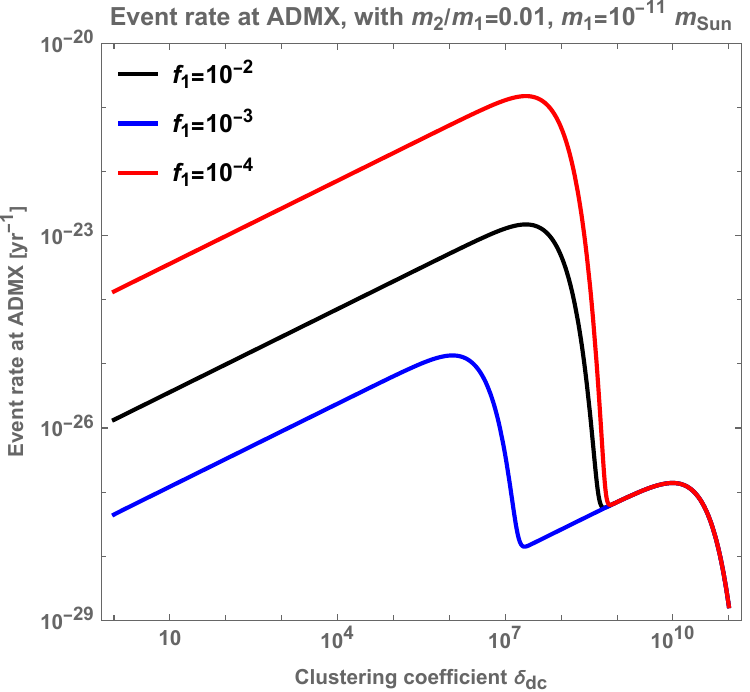}
    }
    \caption{The dependence of the event rate per year on the clustering coefficient $\delta_{\rm dc}$, for three different mass ratios $m_2/m_1=1,\ 10,\ 0.1$ with $f_1=0.01$ and $m_1=10^{-11}\ m_{\rm Sun}$ (left) and for three different values of $f_1=10^{-2},\ 10^{-3},\ 10^{-4}$}
    \label{fig:rates3}
\end{figure}
Fig.~\ref{fig:rates3} studies the effect of clustering, but for a dichromatic mass function, varying a number of parameters, and as a function of the clustering parameter $\delta_{\rm dc}$. Both panels assume $m_1=10^{-11}m_{\rm Sun}$; in the left panel we fix $f_1=0.01$ and vary the mass ratio $m_2/m_1=0.1,\ 1,\ 10$; in the right panel we fix, instead, $m_2/m_1=0.01$, and vary $f_1=0.01,\ 0.001$, and $10^{-4}$. As usual, $f_2$ is maximized for a given $f_1$, as per Eq.~(\ref{eq:f2max}). The first peak, to the left, corresponds to $m_2$, while the one to the right to $m_1$ in the dichromatic mass function.

The main takeaway point in this figure is that the effect of early clustering strongly depends on the mass ratio for a non-monochromatic mass function. Also, notably, at very large clustering there exists an exponential suppression, deriving from the physical effect that mergers in this case have already occurred at very early times and the binaries, thus, have been exhausted into early mergers.

\begin{figure}[!t]
    \begin{center}
    
    \mbox{\includegraphics[height=7.8cm]{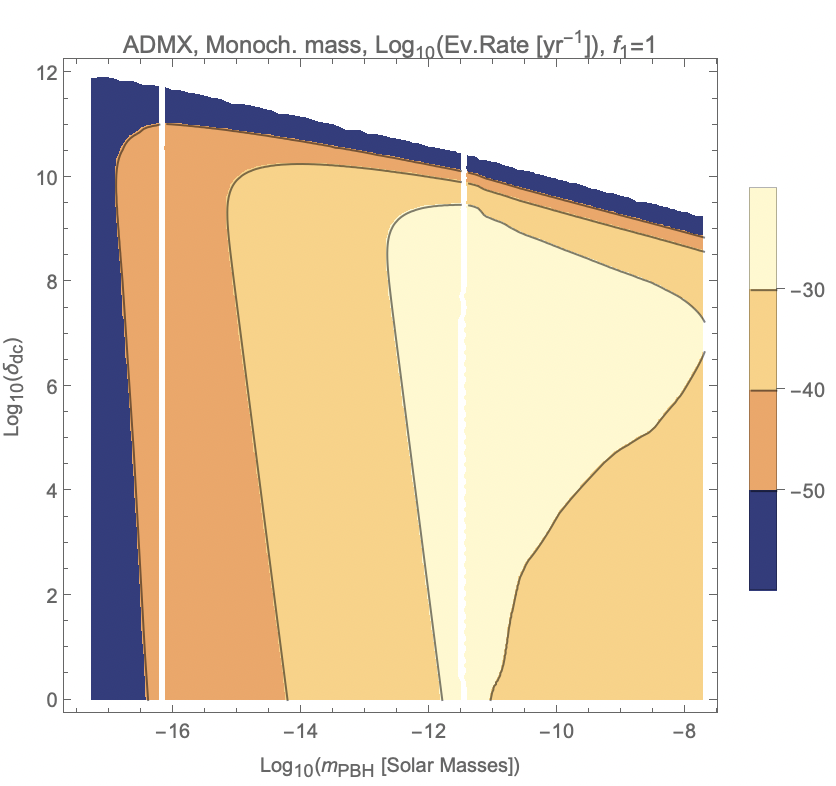}\qquad \includegraphics[height=7.8cm]{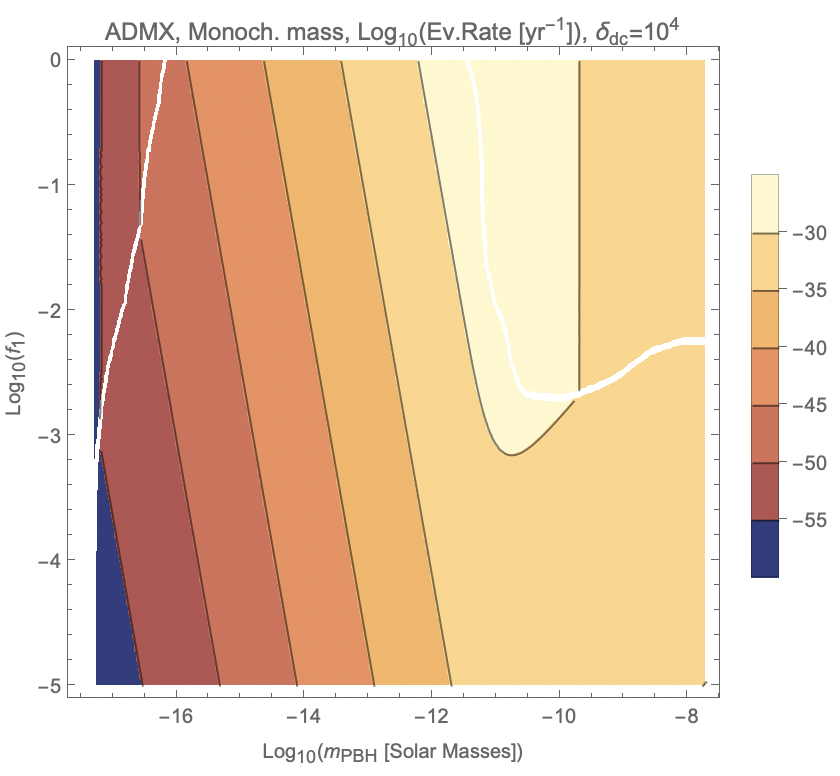}
    } 
    \caption{Contour plots of the Log10 of the event rate per year as a function of the dichromatic mass function ratio $m_2/m_1$, for  $m_1=10^{-12}\ m_{\rm Sun}$, versus $\delta_{\rm dc}$ with $f_1=10^{-5}$ (and maximized $f_2$) (left) and $f_1$ (right) with $\delta_{\rm dc}=10^4$.}
    \label{fig:rates4}
    \end{center}
\end{figure}
The contour plots of fig.~\ref{fig:rates4} illustrate, as above on a Log10 scale, but different color-coding, the event rate on the planes defined by the mass ratio $m_2/m_1$ versus clustering coefficient $\delta_{\rm dc}$ (left) and $f_1$ (right). In both plots we fix $m_1=10^{-12}m_{\rm Sun}$, while we set $f_1=10^{-5}$ in the left panel and $\delta_{\rm dc}=10^4$ in the right panel. Note that as in previous plots, $f_2$ is maximized for a given $f_1$.  The white regions in the left panel have vanishingly small event rates because of the exponential suppression discussed above.

{
The left panel highlights that the maximal rate, approximately independent of $m_2/m_1$, is achieved for $10^1\lesssim\delta_{\rm dc}\lesssim 10^4$, and for the smallest mass ratios $m_2/m_1$. The right panel, instead, shows how the event rate is maximized at extreme mass ratios $m_2/m_1\gg1$ at large $f_1$, or $m_2/m_1\ll1$ at small $f_1$.
}

\begin{figure}[!t]
    \centering
    \mbox{\includegraphics[height=7.5cm]{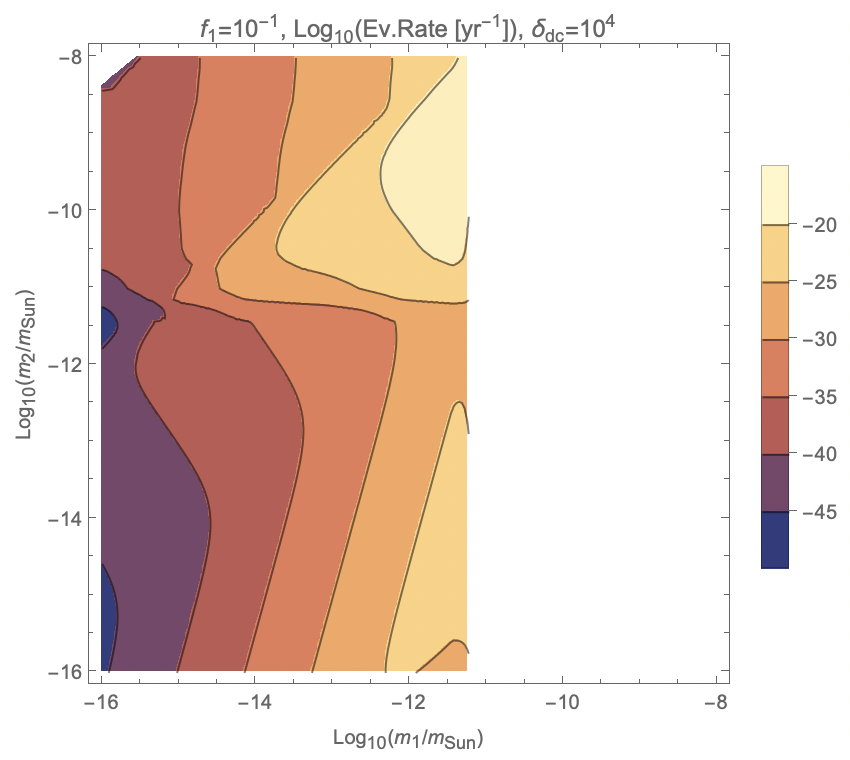}\quad  \includegraphics[height=7.5cm]{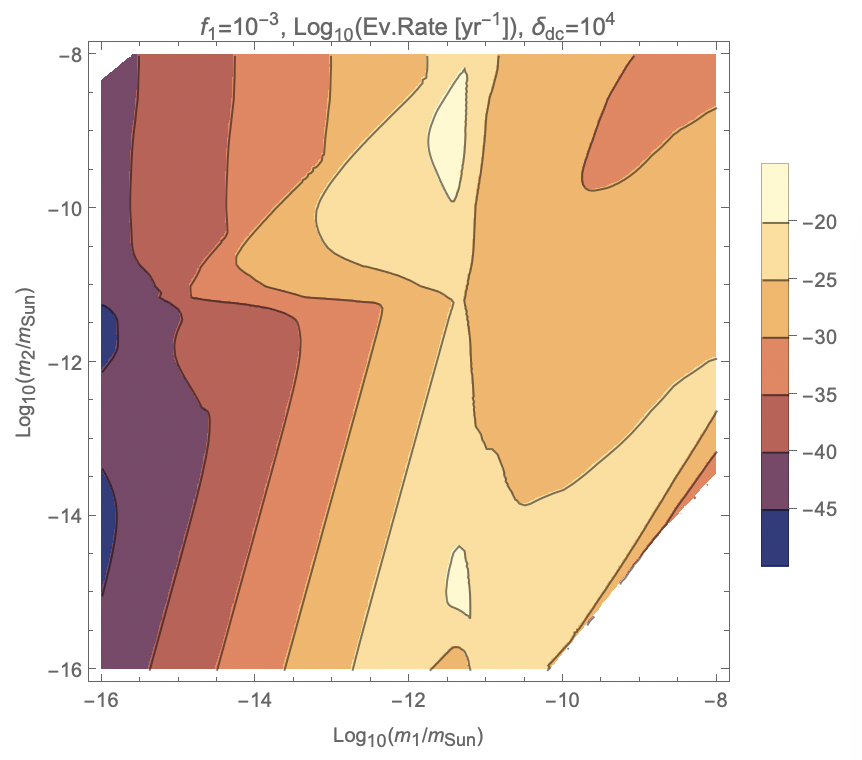}
    }\\[1cm]
       \mbox{\includegraphics[height=7.5cm]{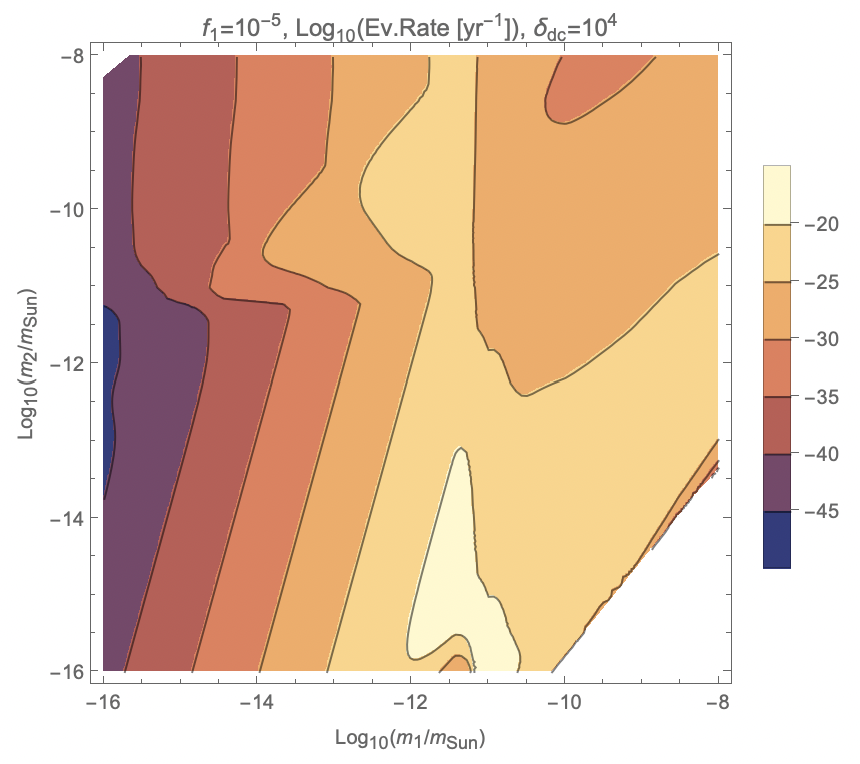}\quad \includegraphics[height=7.5cm]{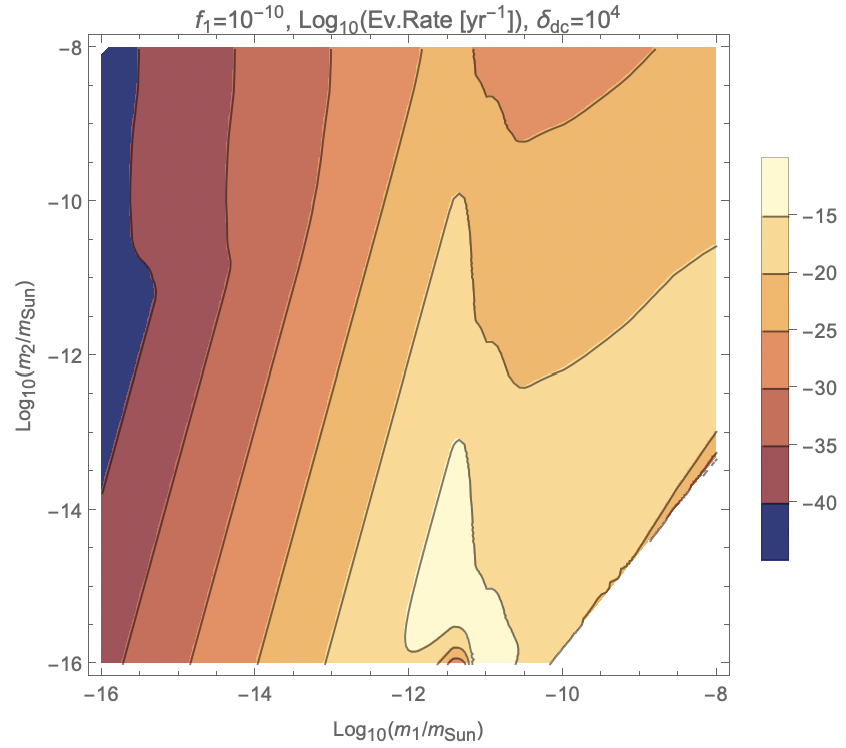}
    }
    \caption{Four panels showing the Log10 contours for the yearly event rate on the $(m_1,m_2)$ plane, all at $\delta_{\rm dc}=10^4$, for $f_1=10^{-1},\ 10^{-3},\ 10^{-5},\ 10^{-10}$ in clockwise order from the top left}
    \label{fig:rates5}
\end{figure}
In the final four panels of fig.~\ref{fig:rates5} we study the event rate, on the plane defined by the  masses in the merging binary $m_1$ and $m_2$. In all panels we fix $\delta_{\rm dc}=10^4$, and, clockwise from the top left, we set $f_1=10^{-1}$, $10^{-3},\ 10^{-5}$ and $10^{-10}$. The white regions correspond to $f_1>f_{\rm PBH}(m_1)$ and are therefore excluded by constraints on the PBH abundance at that mass.

In all cases, we find that rates are largest in the range around $10^{-13}\lesssim m_{1,2}/m_{\rm Sun}\lesssim 10^{-11}$, with peaks at values slightly in excess of $10^{-12}m_{\rm Sun}$. The details of each panel are interesting and straightforward to interpret based on the discussion of the plots above. Note that the color range is not the same for all plots, and that the event rate is increasingly large with decreasing $f_1$.

\section{Comparison of different detectors and frequencies}

We compare, here, the rates expected with improved, future detectors, sensitive to a broader range of frequencies. Specifically, we consider a version of ADMX with a significantly reduced resonant frequency of $\nu=0.65$ GHz (corresponding to  ADMX run1A; note that for instance the recent run1D is around 1.4GHz), with all other parameters set to the default values indicated above. For all detectors we assume 60 seconds integration times.

For the SQMS detector we employed the parameters indicated in Ref.~\cite{Berlin:2021txa}: $f\in (1-2)\ {\rm GHz}$, $Q\sim 10^6$, $|{\mathbf B}|=5$ T, $V_{\rm cav}=100$ L, $T_{\rm sys}=1$ K, and $\eta_n=0.1$ as for ADMX.

For the ADMX-EFR we utilize projections provided by the ADMX Collaboration\footnote{G.P. Carosi, private communication}, featuring a magnetic field – 9.4 T at central field (mean over cavities on the order of 9.1 T), a cavity volume consisting of an 18 cavity array with combined volume of approximately 218 liters in the frequency range 2-3 GHz and 182 liters (3-4 GHz), and a
Cavity Unloaded Quality factor of 120k; the total system noise temperature was assumed to be 440 mK.

Finally - albeit the  geometric structure of the detector may pose non-trivial coupling differences from the linear structure of ADMX - for DM Radio GUT we assume the parameters given in\footnote{https://indico.fnal.gov/event/63051/} and also outlined in \cite{Domcke:2022rgu} (assuming the sensitivity is approximately reflected in our Eq.(\ref{eq:cavity}) above): a magnetic field of 16 T, a volume of 4,580 L, a quality factor of 10$^6$, a system noise temperature of 1 mK, and a frequency of 4 GHz.

\begin{figure}[!t]
    \centering
    \mbox{\includegraphics[height=7.5cm]{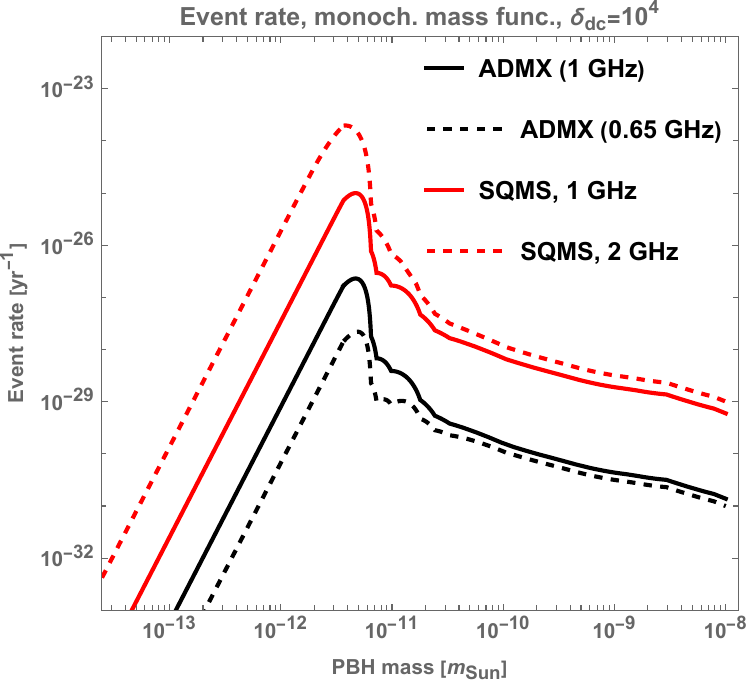}\ \includegraphics[height=7.5cm]{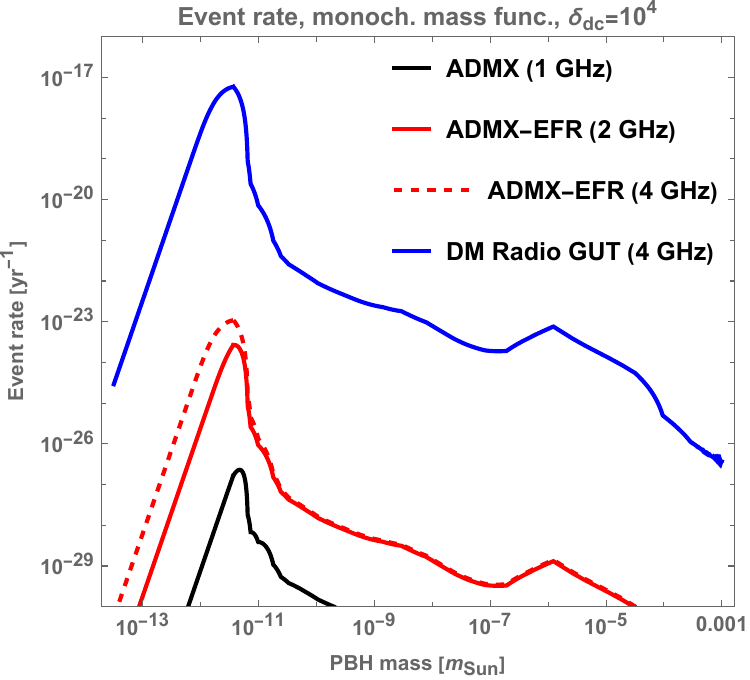}
    }
    \caption{Left: A comparison of the yearly event rate at different frequencies for ADMX (1 and 0.65 GHz) and for SQMS (1 and 2 GHz); Right: The same as in the left panel, but for ADMX-EFR at 2 and 4 GHz, and for DM Radio GUT at 4 GHz. }
    \label{fig:detectors}
\end{figure}

We present a comparison of the expected, maximal yearly event rate with these various setups in fig.~\ref{fig:detectors}. Generally, as expected, higher frequencies of operation extend the detectors sensitivity to lower masses, and increase the range of masses where a significant event rate should occur (see especially the right panel). The scaling of detectors performance in terms of the detector parameters is otherwise clearly evident from the expression in Eq.~(\ref{eq:cavity}).

{ Unfortunately, even in the most optimistic setup under consideration, the event rate at the most promising future resonant cavity setup is found to be below $10^{-17}$ events per year.}


%

\section{Discussion and Conclusions}\label{sec:conclusions}
The realization that a number of devices originally conceived for different experimental purposes could be used in the quest for high-frequency GWs has spurred a resurgent interest in potential signals \cite{Aggarwal:2020olq, Aggarwal:2025noe}. Parallel to this, as the search for weakly-interacting particle dark matter dawns \cite{Arcadi:2017kky, Arcadi:2024ukq} with no compelling evidence of it being realized in nature has also given rise to renewed interest in primordial black holes as dark matter candidates. Interestingly, a broad black hole mass range, between, approximately, masses of $10^{-16}$ and $10^{-10}$ solar masses (the ``asteroid mass range'') is compatible with these objects being the entirety of the cosmological, and Galactic, dark matter \cite{Carr:2017jsz, Carr:2020xqk}.

The asteroid mass window is, however, extremely complicated to probe. At low masses, evaporation via Hawking-Bekenstein radiation is too slow; at high masses microlensing is not effective because of a number of technical reasons related to finite size source effects, wave optics, and the interplay of event duration and observational cadence. However, mergers of such light black holes could give rise to GHz frequency gravitational was signals similar to those detected, at much lower frequency, by the inteferometers LIGO/Virgo/KAGRA, and potentially detectable by the above-mentioned re-purposed detectors.

{Here, we provided an in-depth examination of the expected event rate for light PBH mergers at experiments similar to the microwave resonant cavity ADMX}. Our assumptions are purposely optimistic, stretching the event rates to their maximal possible values. This includes maximizing the effect of clustering upon binary formation, the local density of Galactic dark matter, and the black hole mass function, the latter being customarily and systematically assumed to be monochromatic in previous studies (see e.g. \cite{Franciolini:2022htd, Domcke:2022rgu, Gatti:2024mde}).

{We found that even with future experiments, and with the most-optimistic possible assumptions, the event rate at resonant cavities is vanishingly small, due to the experiments' sight distance being at most a fraction of an AU.}

This study addressed closely the role of a non-monochromatic mass function, showing that event rates can be enhanced by orders of magnitude compared to the usually-assumed monochromatic case; everywhere, we took into consideration the relevant observational constraints on the black hole abundance, as they apply to non-trivial mass functions. We first assessed the most significant pathway to binary formation, and concluded that throughout the parameter space of interest it is associated with early, three-body binary formation. We then studied how event rates depend on a ``dichromatic'' mass function - one that features two subpopulations of different masses in different proportions, which has been shown elsewhere to maximize the merger event rate \cite{Lehmann:2020bby}. We also studied the effect of clustering at early (and late) times, and prospects for current and future detectors alike. { Finally, we demonstrated that the sight distance to black hole merger events is mass-independent for any narrow-band detector; additionally, we demonstrated that the resonant cavity ring-up condition corresponds to the number of cycles spent by the signal within the bandwidth being equal to the cavity's quality factor.}


\section*{Acknowledgements} \label{sec:acknowledgements}
We thank Chelsea Bartram, Nick Du,
Gray Rybka and Gianpaolo Carosi for input, feedback, and discussions. {We are very thankful to Asher Berlin for important feedback on the original version of this manuscript. Finally, we are grateful to the anonymous Referee for prompting us to important discussions of key physics points in a revised version of this manuscript.} This work is partly supported by the U.S.\ Department of Energy grant number de-sc0010107 (SP). 

\appendix
\section{Other merger pathways}

In \cite{Raidal:2024bmm} the authors presented an update of their previous work \cite{Raidal:2017mfl} on early three-body mergers. Albeit an improvement in several aspects of the calculation, this latter discussion does not include a detailed discussion of the effect of early-time clustering. We verified that the differences in the merger rate compared to Ref.~\cite{Raidal:2017mfl} are actually marginal and within a factor of a few, when $\delta_{\rm dc}=1$. As such, we resort to the expression discussed above.

Other merger pathways include the early two-body (E2), late two-body (L2), and late three-body (L3) formation scenarios. For a generic mass function $\psi$, the E2 pathway merger rate for masses $m_1,\ m_2$ reads \cite{Raidal:2024bmm}
\begin{equation}
    \frac{dR_{E2}}{dm_1dm_2}\approx\frac{1.6\times 10^6}{{\rm Gpc}^3\ {\rm yr}}f_{\rm PBH}^{\frac{53}{37}}\eta^{-\frac{34}{37}}\left(-\frac{32}{37}\right)^{-\frac{32}{37}}S_LS_E\frac{\psi(m_1)}{m_1}\frac{\psi(m_2)}{m_2}
\end{equation}
where as usual $f_{\rm PBH}=\int\psi(m)dm$, $\eta=m_1m_2/(m1+m_2)^2$, $M=m_1+m_2$, and $S_L$ and $S_E$ are suppression factors (which for the sake of comparison with the $E3$ rate we set to 1 below).

The late, 2-body pathway leads to a merger rate \cite{Raidal:2024bmm}
\begin{equation}
    \frac{dR_{E2}}{dm_1dm_2}\approx\frac{3.4\times 10^{-6}}{{\rm Gpc}^3\ {\rm yr}}f_{\rm PBH}^{2}\delta_{\rm eff}\left(\frac{\sigma_v}{\rm km/s}\right)^{-\frac{11}{7}}\eta^{-\frac{5}{7}}\frac{\psi(m_1)}{m_1}\frac{\psi(m_2)}{m_2},
\end{equation}
where $\delta_{\rm eff}$ is a local DM density contrast factor, and we assume a local velocity dispersion $\sigma_v\sim 100$ km/s.

Finally, the 3-body late mergers reads \cite{Raidal:2024bmm}
\begin{eqnarray}
    \nonumber\frac{dR_{L3}}{dm_1dm_2}&\approx&\frac{1.3\times 10^{-16}e^{-6.0(\gamma-1)}}{{\rm Gpc}^3\ {\rm yr}}f_{\rm PBH}^{3}\delta^2_{\rm eff}\left(\frac{\sigma_v}{\rm km/s}\right)^{-9+\frac{8\gamma}{7}}\eta^{-1+\frac{\gamma}{7}}{\tilde{\cal F}}\left(\frac{\langle m\rangle}{2\eta M}\kappa_{\rm min}\right)\frac{\psi(m_1)}{m_1}\frac{\psi(m_2)}{m_2},
\end{eqnarray}
where the function ${\tilde{\cal F}}$ is detailed in Ref.~\cite{Raidal:2024bmm}, and $1<\gamma<2$, which determines the initial angular momentum distribution of the binaries,
$$
\frac{dP}{dj}\sim \gamma j^{\gamma-1},
$$
is assumed to be 1.1 (we find that the induced uncertainty from varying $\gamma$ is mild).

\begin{figure}[!t]
    \centering
    \mbox{\includegraphics[height=6.3cm]{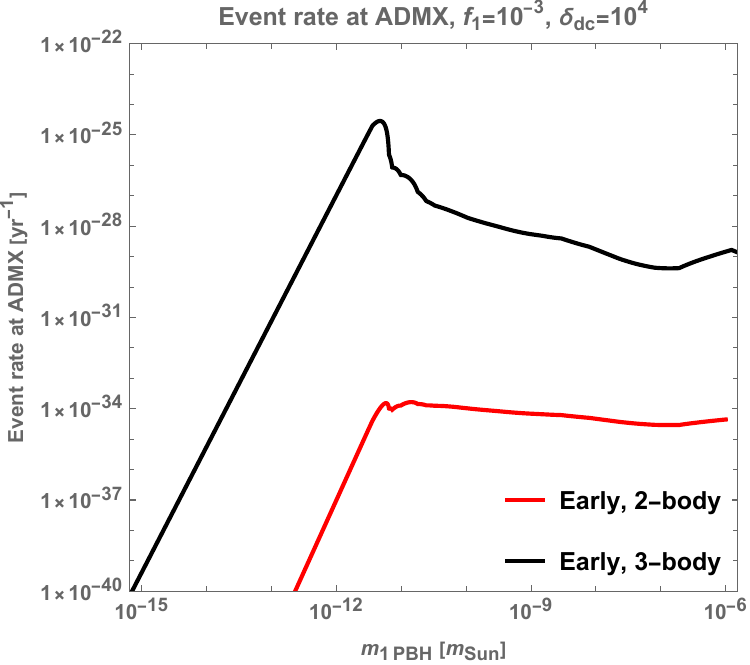}\qquad \includegraphics[height=6.3cm]{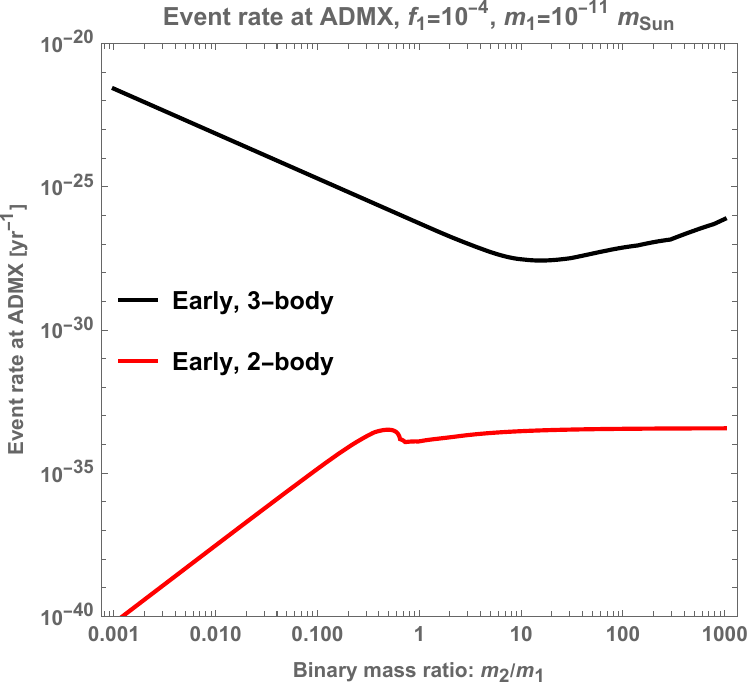}
    }
    \caption{Pathways comparison -- see the text for details.}
    \label{fig:pathways}
\end{figure}

We compare the four pathways first, for simplicity, using a monochromatic mass function in the left panel of Fig.~\ref{fig:pathways}. We assume a density contrast $\delta_{\rm dc}=10^4$ for all pathways (we implement it as described at the end of Ref.~\cite{Raidal:2024bmm} for the pathways E2, L2 and L3) and we use for each mass the maximal possible value of $f_{\rm PBH}(m_{\rm PBH}$ for a given mass. The figure shows that the E3 pathway dominates everywhere between 6 and 7 orders of magnitude to the next most significant pathway.  { The late 2- and 3-body pathways are several orders of magnitude below the range shown in the figure.}

The right panel of fig.~\ref{fig:pathways} shows the results for a dichromatic mass function, with the mass function mass ratio $m_2/m_1$ shown in the $x$ axis. Here again it is fully apparent how the E3 pathway dominates other pathways by several orders of magnitude, justifying in full the approach we take in this analysis. { Again here, the late 2- and 3-body pathways are several orders of magnitude below the range shown in the figure.

\section{Clustering Effects on Black Hole Binary Formation Pathways}

PBH clustering plays a dual and complex role in influencing binary formation pathways, with distinct effects on early and late binary formation channels. Clustering suppresses early 3-body binary formation in a radiation-dominated Universe. This suppression occurs due to perturbations induced by nearby PBHs, tidal forces, angular momentum exchanges, and hierarchical dynamics, which destabilize nascent binaries and reduce merger rates. Studies consistently show that clustering significantly lowers early binary merger rates, particularly at high PBH dark matter fractions, with nearest-neighbor distances playing a critical role in suppressing binaries \citep{raidal2024, escriva2022, luca2020, jedamzik2020, ballesteros2018}. However, a subset of early binaries may survive in regions of low clustering density or when initial conditions are favorable to stable configurations \citep{luca2020}. Studies find that, generally, such effects amount to a rate suppression by at most two-three orders of magnitude \citep{raidal2024, escriva2022, luca2020, jedamzik2020, ballesteros2018}.

In contrast, PBH clustering enhances late binary formation, particularly through 2-body dynamical capture within dense environments such as dark matter halos or small PBH clusters. These late pathways arise from increased interaction rates, clustering-driven density amplification, and gravitational harmonics that promote orbital binding. Studies highlight the role of nested overdensities and evolving structure formation in facilitating such interactions and highlight the resulting high-eccentricity late binaries detectable through gravitational wave signals \citep{raidal2024, stasenko2024, luca2020, delos2024, aljaf2024}. While this enhancement boosts late-time merger rates, competing effects like cluster heating and evaporation in dense PBH clusters could partially inhibit late binary formation in some scenarios \citep{korol2019}. In the present case, late-time merger rates are suppressed by over 30 orders of magnitude, so we do not expect the hieararchy of merger pathways we assume here to change.

Clustering's influence evolves over cosmic time, transitioning from suppression of early (post-recombination) binaries to amplification of late-stage binary formation during structure formation epochs dominated by dark matter halos. Redshift-dependent analyses indicate that suppression mechanisms are most prominent at high redshift, while clustering-enhanced late pathways dominate in lower redshift environments \citep{raidal2024, stasenko2024, luca2020, aljaf2024}. This dual role of clustering raises important questions about the exact thresholds and transitions between suppression and enhancement mechanisms, as well as the downstream effects of disrupted early binaries on late-time formation pathways \citep{raidal2024, stasenko2024, luca2020, delos2024}.


\section{Microlensing and Gravitational Wave Constraints}

PBHs within the mass range of Earth mass to \(10^{-16} M_\odot\) present unique challenges and opportunities for observational detection, primarily through microlensing and GWs methods. Microlensing surveys, such as OGLE and Subaru/HSC, have robustly constrained Earth-mass PBHs, with strong restrictions imposed on their dark matter fraction. However, for smaller masses (below \(10^{-12} M_\odot\)), finite source size effects significantly reduce microlensing sensitivity. Femtolensing of gamma-ray bursts can, in principle, probe PBHs as small as \(10^{-16} M_\odot\), but this method's efficacy is limited by the extended size of gamma-ray burst sources, calling into question earlier constraints \cite{niikura2019constraints, katz2018femtolensing, pfrang2021optical, petac2022microlensing}. Additionally, clustered PBHs marginally impact microlensing event rates, as compactness requirements for clustering are often unmet, rendering practical deviations insignificant in most models \cite{toshchenko2019microlensing, petac2022microlensing}.

Continuous gravitational wave (GW) detection methods show promise for identifying binaries of planetary- and asteroid-mass PBHs, especially for masses in the \(10^{-12} M_\odot\) to \(10^{-5} M_\odot\) range. Studies using data from LIGO/Virgo’s O3a observing run, such as those by Miller et al., have constrained PBH binaries through continuous wave (CW) searches, targeting signals from tightly bound compact binaries. While transient GW signals from mergers dominate for larger masses, continuous and transient continuous-wave (tCW) methods are more relevant for PBHs in this lower mass regime. Nevertheless, sub-\(10^{-12} M_\odot\) PBHs remain largely undetectable with current GW observatories due to sensitivity limitations and the lack of strain signatures in the accessible frequency range. Future high-frequency GW detectors like ADMX and space-based observatories like LISA may be able to probe these smaller masses in clustered or binary configurations \cite{miller2021constraints, miller2024gravitational, domenech2024probing}.

PBH clustering and binary formation are central to enhancing GW detection prospects, as clustering increases the likelihood of binary interactions and GW emission. While clustering may not substantially affect microlensing constraints, it has been proposed as a mechanism to boost GW merger rates, particularly for asteroid-mass PBHs forming in dense environments, as noted and explored above. Despite these theoretical advancements, the observational integration of microlensing and GW detection methods remains underdeveloped. Overlapping sensitivity windows (e.g., microlensing at Earth mass and GW constraints at planetary mass) could jointly refine PBH abundance and clustering properties. However, the challenges introduced by finite source size effects, computational limitations in GW template searches, and uncertainties in clustered PBH models hinder the realization of such synergies \cite{toshchenko2019microlensing, miller2021constraints, miller2024gravitational, niikura2019constraints}.

In summary, microlensing and GW methods offer complementary, but not entirely overlapping, approaches to constraining PBHs across this mass range. While microlensing excels at Earth-mass PBH detection, its sensitivity diminishes at the lower end of the spectrum. In contrast, GW techniques are increasingly effective for clustered and binary systems in the planetary-mass regime but remain largely impractical for detecting individual small PBHs below \(10^{-12} M_\odot\). To respond to these challenges, future research must refine clustering models, consider next-generation GW facilities, and develop joint observational frameworks to bridge the gaps in PBH detectability \cite{miller2024gravitational, katz2018femtolensing, petac2022microlensing}.

}




\bibliographystyle{unsrt} 
\bibliography{bib}

\end{document}